# Reconfigurable Free-Space Optical Interconnects

Jaekwon Lee[1,2], Carson G. Valdez[1,2], Anne R. Kroo[1,2], David A. B. Miller[1,2], and Olav Solgaard[1,2,*]

[1]Edward L. Ginzton Laboratory, Stanford University, 348 Via Pueblo, Stanford, 94305, CA, USA

[2] Electrical Engineering, Stanford University, 350 Jane Stanford Way, Stanford, 94305, CA, USA

[*]Corresponding author. E-mail: solgaard@stanford.edu

## Abstract

Free-space optical interconnects can support many low-data-rate channels while avoiding the cost, energy consumption, and packaging constraints of optical waveguide- or fiber-based systems that are widely used today. We propose a free-space optical interconnect system based on reconfigurable micromirror arrays, and show that it scales favorably due to its two-dimensional scaling. We discuss the channel count scaling model, compare our demonstration with theory, and project a channel count above 10,000 to achieve 100 Tbps with slow channels (e.g., <10 Gbps). In a proof-of-concept system with a piston-motion micromirror array, we demonstrate nine channels that are simultaneously optimized. The optimization enables -21.7 dB of average worst-case crosstalk for a channel pitch equal to twice the beam radius, a smaller channel spacing than that required in flat tip-tilt mirrors. The optimization flow is further investigated for different channel spacings, simultaneous illumination, and different numbers of outputs, which generally achieve smaller worst-case crosstalk than the flat tip-tilt mirror case. Overall, the work demonstrates the advantage of piston-motion micromirrors in the proposed optical interconnects.

Short-haul optical interconnects have emerged as a promising solution for reducing datacenter energy consumption[1], provided that they can simultaneously achieve <1 pJ/bit, >100 Tbps/socket, and <$0.10/Gbps[2]. Reaching these goals with waveguide- or fiber-based systems has proven challenging[2–6]. Single-mode optical fibers, fiber array units, and other fiber-related accessories together account for > 30 % of the co-packaged optics (CPO) switch cost[6], even with limited deployment of a fiber count of approximately 1,000. Thus, large bandwidth per channel (i.e., >200 Gbps/lane) is necessary, introducing a trade-off between energy consumption and total bandwidth. On the other hand, using energy-efficient slow fiber channels (i.e., <10 Gbps/channel) is constrained to a total achievable bandwidth of 1-5 Tbps due to the fiber count limit[5,7].

Thus, numerous slow channels approach (i.e., slow-and-wide system) requires reliable fiber-free communication[8,9], with >10,000 channels to achieve 100 Tbps with ~10 Gbps channels[10]. With the advent of energy-efficient but slower light sources such as micro-light-emitting diodes (micro-LEDs)[5], systems that scale beyond 30,000 channels might be beneficial. Micromirror-based free-space links have been proposed by a few studies utilizing vertical-cavity surface-emitting lasers (VCSELs) and tip-tilt motion micromirrors[11,12]. However, the demonstration was performed without actual reconfigurability[11] or with only one-dimensional reconfigurability with complex modulation mechanisms[12], which is not suitable for slow-and-wide systems. On the other hand, fiber-optic switches[13,14] show scalable demonstrations, but they focus on fiber-based systems and thus were limited to ≤300 channels with low fill-factor mirrors without further scalability to >1,000 channels.

In this work, we propose a free-space optical interconnect system formed by a pair of micro-electromechanical systems (MEMS) micromirror arrays as two-dimensional reconfigurable photonic channels for surface-emitting light sources such as VCSELs or micro-LEDs. We demonstrate a proof-of-concept experiment that shows two-dimensional reconfigurability of the proposed system, which potentially scales to >1,000 channels with the current technology. Our results also suggest a roadmap for further channel count scaling.

## Scaling of Free-Space Optical Interconnects

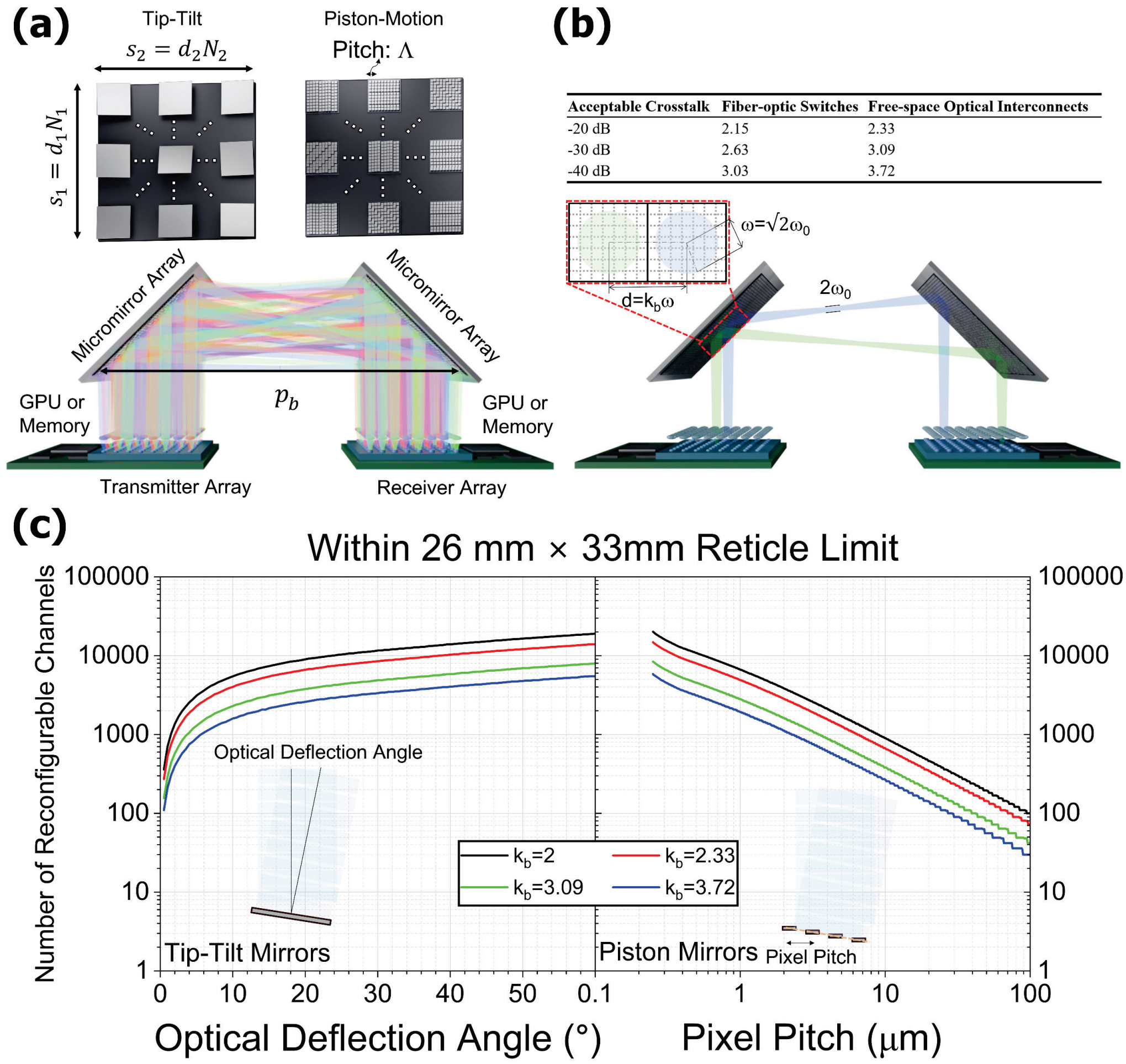

**Fig. 1 | Proposed Free-space optical interconnects.** (a) Schematic of the system. Elements in the transmitter and receiver arrays are mapped by a pair of micromirror arrays, enabling vertical incidence to the chips. The free-space nature of the system enables space-efficient beam overlap. The figure is shown for a 10×10 array of communication channels. (b) Description of the crosstalk structure. The channel pitch is defined as $k_b$ times the beam radius ω at the mirror, and the worst-case crosstalk is strongly dependent on $k_b$. (c) The scaling of reconfigurable channel counts within a reticle limit for both tip-tilt and piston mirror arrays, as a function of micromirror parameters: maximum optical deflection angle for tip-tilt mirrors, and smallest pixel pitch for piston-motion mirrors. Larger reticles will enable larger channel counts, proportional to the square root of the reticle area scaling.

Figure 1 (a) illustrates the proposed system, utilizing microlens-integrated surface-emitting light sources, two micromirror arrays, and microlens-integrated photodetector arrays[15]. The system can utilize either tip-tilt mirrors that geometrically deflect the beam through mechanical rotation in two different directions[14], or piston-motion mirrors that collectively impose a phase tilt on the reflected beam[16]. In tip-tilt architectures, each channel is supported by two tip-tilt mirrors: one that directs one input beam onto one output mirror that, in turn, directs that beam onto the desired output channel. Piston-mirror architectures also use one input and one output, but here the reflectors consist of multiple piston mirrors that deflect the beam through diffraction. The two micromirrors are packaged at 45° from the chip plane, which is the optimal angle (Supplementary Information). The system has several benefits over existing fiber-optic switches[13,14] — (1) short wavelengths, λ=450 nm or shorter, improve channel density; (2) packaging cost is reduced as the channels can be aligned *in situ* instead of fine-alignment packaging; (3) the fabrication cost is governed by a parallel lithography process instead of serial fiber drawing; and (4) all input/output beams remain vertical to the chip for energy-efficient (i.e., miniaturized) transmitters and detectors[8].

Although scaling studies of fiber-optic switches can be found in the literature[13,17], we suggest an alternative set of equations that remove the three main assumptions to extract a more accurate channel count—(1) the maximum optical deflection angle is small (small-angle approximation); (2) the separation between any mirrors is constant (constant-separation approximation); and (3) the chip footprint is square (square-chip approximation). The equation is still governed by the trade-off between aperture size and separation (i.e., diffraction) as illustrated in Supplementary Information:

$$s_1 = \frac{N_1(N_1-1)\lambda k_b^2}{\pi cos\beta}\frac{1+tan\alpha_1}{tan\alpha_1}\ (0<\alpha_1<\frac{\pi}{4}) \tag{1a}$$

$$s_2 = \frac{N_2(N_2-1)\lambda k_b^2}{\pi tan\alpha_2}\frac{1}{1-\frac{2(N_1-1)d_1 sin\beta}{p_b}} \tag{1b}$$

where $k_b$ is an application-dependent parameter indicating channel spacing (i.e., the channel pitch is defined as $k_b$ times the beam radius ω at the mirror), λ is the wavelength, β=45° is the incidence angle of the beam, $p_b$ is the longest linear separation between the two arrays, $s_1(s_2)=N_1d_1$ ($N_2d_2$), $N_1(N_2)$, $\alpha_1$ ($\alpha_2$), and $d_1(d_2)$ are the minimum lengths of the chip, linear counts of channels, maximum optical deflection angles, and aperture sizes along each axis, respectively.

The same equations hold for piston-motion mirrors, while the optical deflection angle is practically $\alpha_1 = \sin^{-1}(\lambda/2\Lambda\cos\beta)$ and $\alpha_2 = \sin^{-1}(\lambda/2\Lambda)$[16,18], where Λ is the pitch of the piston mirror array:

$$s_1 = \frac{N_1(N_1-1)\lambda k_b^2}{\pi cos\beta}\frac{1+\tan\{\sin^{-1}(\lambda/2\Lambda\cos\beta)\}}{\tan\{\sin^{-1}(\lambda/2\Lambda\cos\beta)\}} \approx \frac{2N_1(N_1-1)k_b^2\Lambda}{\pi} \tag{2a}$$

$$s_2 = \frac{N_2(N_2-1)\lambda k_b^2}{\pi\tan\{\sin^{-1}(\lambda/2\Lambda)\}}\frac{1}{1-\frac{2(N_1-1)d_1 sin\beta}{p_b}} \approx \frac{2N_2(N_2-1)k_b^2\Lambda}{\pi} \tag{2b}$$

The simplified forms come from the small-angle approximation. Each channel consists of $M_1 \times M_2$ piston-motion mirrors:

$$M_1 = \frac{s_1}{N_1\Lambda} \approx \frac{2(N_1-1)k_b^2}{\pi} \tag{3a}$$

$$M_2 = \frac{s_2}{N_2\Lambda} \approx \frac{2(N_2-1)k_b^2}{\pi} \tag{3b}$$

indicating that the number of piston-motion mirrors per channel roughly scales linearly with the channel count. For instance, 100 channels require each reflector to be an array of 23×23 mirrors, assuming $k_b$=2.

Figure 1 (b) shows the role of $k_b$ in Eq. (1a-3b). Unlike conventional fiber-optic switches, the light vertically incident on the microlens can be coupled into the photodetector regardless of the mode overlap. Therefore, the worst-case crosstalk is calculated as the power ratio captured by an aperture:

$$Crosstalk = 10\log\left[\frac{\int_{-\frac{d}{2}}^{\frac{d}{2}}\int_{-\frac{d}{2}}^{\frac{d}{2}}\exp\left\{\frac{-2\{(x-d)^2+y^2\}}{\omega^2}\right\}dx\,dy}{\int_{-\frac{d}{2}}^{\frac{d}{2}}\int_{-\frac{d}{2}}^{\frac{d}{2}}\exp\left\{-\frac{2(x^2+y^2)}{\omega^2}\right\}dx\,dy}\right] \tag{4}$$

where d=$k_b$ω is the channel size and ω is the beam radius at the mirror. Figure 1 (b) inset presents the $k_b$ values for -20 dB (2.15 vs 2.33), -30 dB (2.63 vs 3.09), and -40 dB (3.03 vs 3.72) of theoretical crosstalk for both fiber-optic switches and the proposed system. It is worth noting that Eq. (4) assumes flat mirrors and fundamental Gaussian beams. In real systems, the required $k_b$ is often larger than this prediction (e.g., $k_b$≈3.5 for -40 dB[13]) due to non-idealities.

The optimal channel counts can be found by maximizing the integer $N_1N_2$ under the bounds of Eq. (1a-1b) or Eq. (2a-2b) with the circular beam assumption $d_1 cos\beta = d_2$ at a given α (Λ) and $k_b$. Figure 1 (c) shows the channel-count scaling. The number of channels is limited by the standard reticle size of 26 ×33 mm$^2$, and in the case of a larger reticle size, the number of channels scales with the square root of the reticle area. A larger α (smaller Λ) minimizes the separation and thus diffraction, while smaller $k_b$ maximizes the channel density at given diffraction limits, both increasing the channel density. The proposed system scales better than the waveguide-based approach (e.g., 80-128 channels within reticle limit[3]) and the fiber-based approach (e.g., 136-300 channels per package[14]). Furthermore, the proposed approach makes multi-reticle approaches more practical, minimizing stitching errors. For instance, ≈30,000 channels can be achieved by aligning three separate 10,000-channel reticles. Still, scaling to >10,000 channels can be challenging: for -20 dB crosstalk, it requires a mechanical deflection angle of ≥19.25° (i.e., optical deflection angle of ≥38.5°) or a piston mirror pitch of ≤360 nm. However, the anisotropic analysis here reveals that angles required for the tilted axis are significantly smaller than the other axis— the example above requires 8.6° mechanical angle or the pitch of ≤ 1.08 μm along the tilted axis. This is favorable for micromirrors, as the former relaxes the spring design for tip-tilt mirrors and the latter ensures larger space beneath the mirrors for transistor integration. Overall, the proposed system shows a clear path towards >10,000-channel scaling.

## Experimental Setup

Figure 2 (a) illustrates the experimental setup. Light from a 450 nm laser diode is reflected at the micromirror array at a nearly 45° angle of incidence. The Texas Instruments phase-light modulator (TI-PLM)[16] with 10.8 μm pitch, 94.5 % fill factor, and 4-bit depth (i.e., 16 phase levels)[16] is controlled by the display interface[19]. A fixed mirror reflects the light back to the TI-PLM for a second reflection and a CMOS camera is used to replace the microlens-integrated photodetector array, emulating the system in Figure 1 (a-b). The beam is focused at the fixed mirror and the TI-PLM is approximately one Rayleigh length away from the fixed mirror[17]. The beam at the TI-PLM has the beam radii of 377 μm × 332 μm, which is projected onto the mirror as 533 μm × 332 μm. The phase levels of the TI-PLM are recalibrated to resolve the reduction of the effective stroke by $\sqrt{2}$ due to oblique incidence (Supplementary Information).

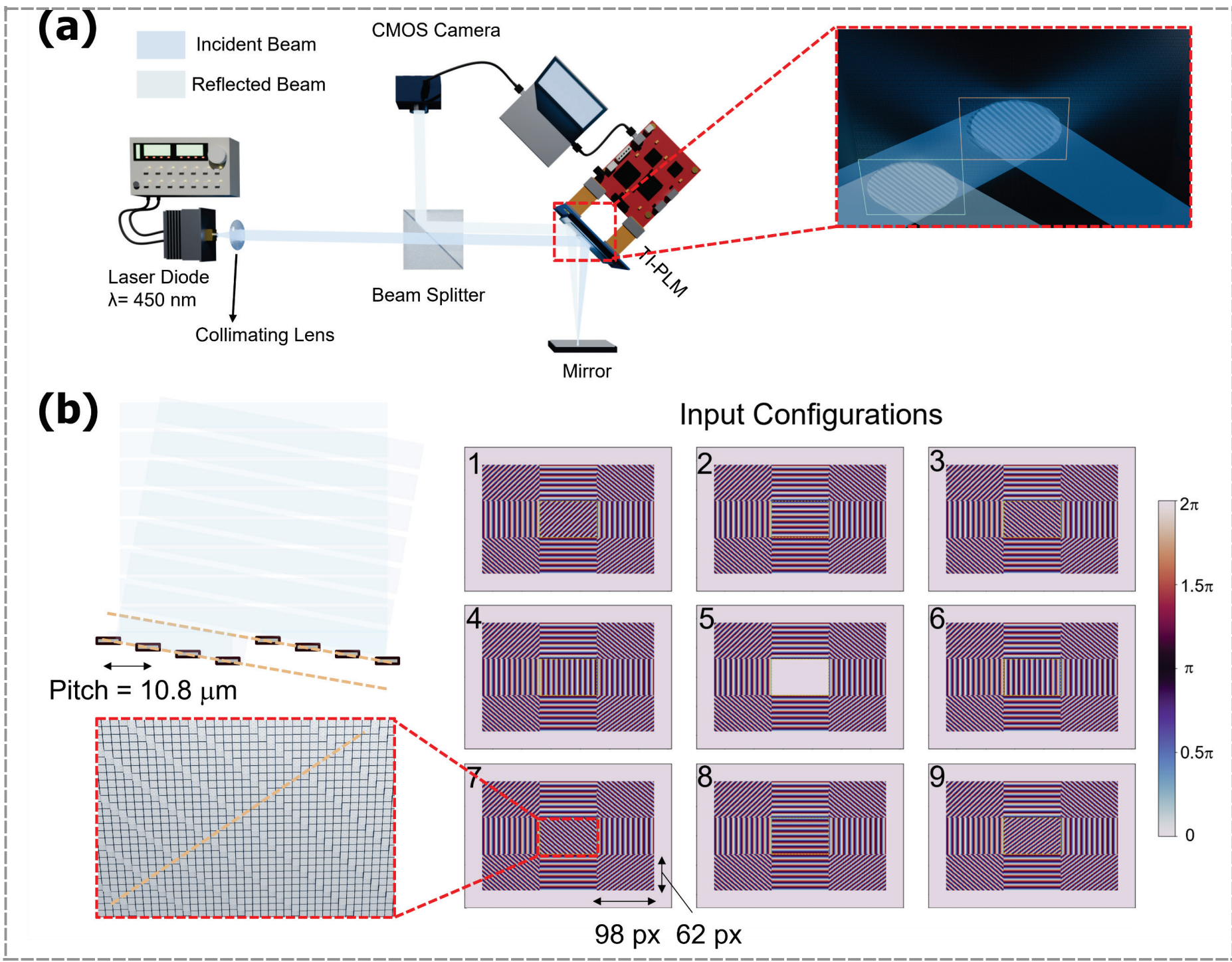

**Fig. 2 | Experimental setup.** (a) The experimental setup aims to emulate the proposed device in Figure 1 (a) and evaluate its performance. A laser diode emitting at 450 nm and focused by a lens is used as a transmitter array

element. Each region of interest in the CMOS camera corresponds to a receiver array element with a microlens and photodetector. A fixed mirror makes the beam reflect twice off different regions of the TI-PLM, equivalent to a pair of micromirrors. The CMOS camera output and TI-PLM input are connected to a computer, making an optimization loop. (b) Nine input configurations were evaluated before optimization, with initial settings set as phase-gradient patterns to effectively create tip-tilt beams.

## Results

Figure 3 (a) shows the initial settings of the micromirror array as nine (i.e., 3×3) different reflectors, each measuring 98 × 62 pixels, corresponding to $k_b$=2. Figure 2 (b) shows nine configurations utilizing these reflectors. In configuration *i*, the tip-tilt (i.e., phase-ramp) pattern first steers the beam from the center reflector to the desired reflector *i* (e.g., *i*=5 for flat pattern), and then reflector *i* cancels out the returning beam angle by holding a 180°-rotated copy of the center pattern. Only the center reflector changes between configurations, and the surrounding eight reflectors are shared across all nine configurations. This emulates a densely packed array where surrounding reflectors are set for their own active channels. Thus, a pattern from one configuration affects the other configurations as undesirable crosstalk, where the beam not captured by the main reflector (e.g., ~9 % of the initial power) is captured by surrounding reflectors and collected in undesired channels, resulting in -16.2 dB crosstalk even for an ideal Gaussian beam. Figure 3 (a) shows that calculated tip-tilt patterns on the piston mirror array (i.e., before the optimization) introduced much larger average worst-case crosstalk of -4.7 dB due to the non-ideal beam profile, non-unity fill factor of the array, and limited phase resolution. After the initial setup, all nine channels are co-optimized with the flow designed for photodetector arrays, as described in Methods. Figure 3 (b) shows the optimization results. The system after the optimization performed better than the ideal value of -16.2 dB, with the average worst-case crosstalk of -21.7 dB. Figure 3 (c) shows the optimization of the maximum crosstalk from ~ -3 dB to < -20 dB, and Figure 3 (d) evaluates the integrated intensity of each channel. Comparing Figures 3 (c) and (d), the signal loss becomes dominant when the optimization moves to fine pixels due to the scattering loss of the complex patterns and the limited bit depth of TI-PLM. For instance, channel 4 experiences three dips in the signal loss, all corresponding to transitions to 12×12, 24×24, and 36×36 gradients, respectively. Channel 7 is affected the least with -0.25 dB of signal loss, while the absolute intensity is always the highest in channel 5, which started from a flat phase level 0. The average loss due to this optimization was -1.24 dB, which will be improved with better phase resolution[20]. Figure 3 (e) shows the crosstalk map before and after the optimization. After the optimization, the crosstalk mostly lies in the -30 to -20 dB range, which is not achievable with flat tip-tilt mirrors at $k_b$=2. Thus, the optimization corrected both the non-ideal beam profile and experimental nonidealities, which can also be used for inexpensive multimode VCSELs or micro-LEDs as light sources. The feasibility of utilizing multimode sources is described in the Supplementary Information.

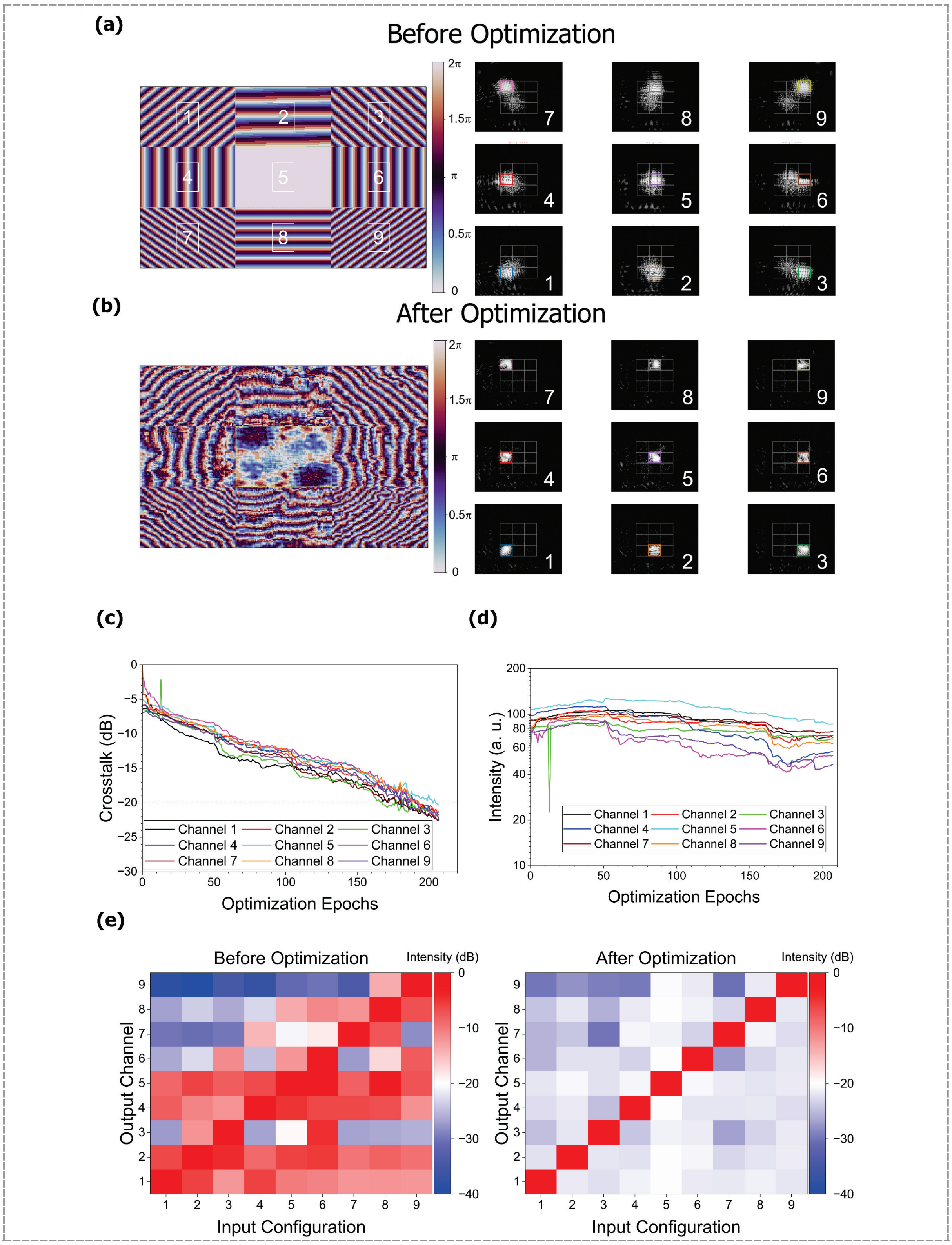

**Fig. 3 | Optimization of nine different channels.** (a-b) Evaluation of nine channels with (a) tip-tilt phase gradient pattern only, before optimization, and (b) pixel-by-pixel optimization. Nine configurations are tested, where each configuration occupies one main channel and the crosstalk to other channels are evaluated. The optimization greatly reduces the crosstalk and maintains a well-defined output to each channel. (c-d) The optimization trade-off. (c) the crosstalk, and (d) the signal loss. Signal loss occurs when the optimization enters a finer stage, which is caused by discrete phase levels and resultant scattering loss between pixels. (e) Crosstalk evaluation before and after optimization. Blue-hue data points indicate crosstalk less than -20 dB, and red-hue data points indicate crosstalk greater than -20 dB. The optimization suppresses all crosstalk to < -20 dB.

Figure 4 compares the optimization for $k_b$=2, 2.33, and 3.09, where the theoretical crosstalk is -16.2 dB, -20 dB, -30 dB, respectively. Figure 4 (a) compares the optimization for different $k_b$ values. Larger $k_b$ values produced smaller crosstalk at the same epoch. The final crosstalk was -21.7 dB for $k_b$=2, -24.3 dB for $k_b$=2.33, and -24.5 dB for $k_b$=3.09. Meanwhile, the average signal intensity starts from a lower value as $k_b$ increases, which is the combined effect of (i) larger ROI, resulting in a lower signal-to-noise ratio, and (ii) lower beam steering efficiency for larger beam steering angles[21]. As a result, the measured crosstalk was limited by the camera noise floor for $k_b \geq 2.33$, as shown in Figure 4 (b). With refined hardware such as low-noise photodetectors, lower crosstalk is expected: -26.4 dB for $k_b$=2.33 and -32.0 dB for $k_b$=3.09.

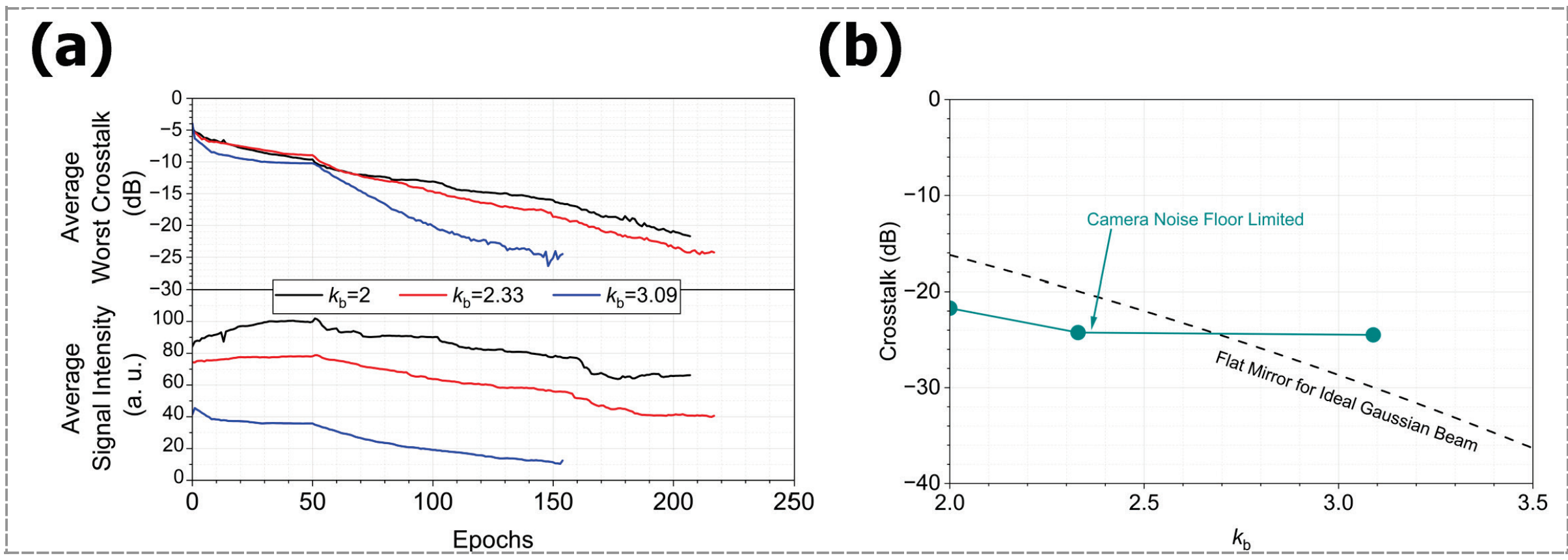

**Fig. 4 | Optimized Crosstalk as a function of $k_b$ (2.0,2.33, and 3.09).** (a) Comparison of the average worst crosstalk and average signal intensity optimization for all three cases. The average worst crosstalk converges faster for larger $k_b$, but the average signal intensity also begins with a lower number and further decreases with optimization for larger $k_b$. (b) The optimization result is compared with the theoretical value for flat mirrors. Optimized piston mirrors achieve better crosstalk until $k_b$=2.33, then limited by the camera noise floor.

In real systems, each channel will be illuminated by an individual light source and modulated separately (e.g., > GHz). To show the feasibility of optimization in real systems, Figure 5 (a) shows the optimization of multiple one-to-one channels under simultaneous illumination. Nine channels are illuminated by a quasi-flat-top beam that covers all the channels, and each channel is modulated between the ON state (i.e., target pattern) and OFF state (i.e., binary grating with phase levels 6 and 13, directing the light out of the beam path) from the mirror pattern itself. Figure 5 (a) shows that the optimization can be performed in the presence of other channels' light by modulating each channel individually, as shown in Figure 5 (b). During the optimization, the average OFF/ON ratio is optimized from -8.33 dB to -17.02 dB, below the ideal value of -10.1 dB at $k_b$=2 that considers the effect from all nearby channels. This experiment shows a non-ideal case and is thus not the limit of the system. First, the quasi-flat-top beam represents the non-ideal case that degrades the crosstalk performance. Second, the modulation speed of the mirrors and the capture rate of the CMOS camera limit the rate to <1 Hz. Finally, the modulation state is not perfect with a large background, because phase differences between phase levels 6 and 13 were not exactly π, leading to a leak in the $0^{th}$ order diffraction.

Figure 5 (c) compares the theoretical calculations with our demonstration. Piston-motion mirrors offer an advantage in channel scaling by enabling $k_b$=2 with crosstalk of <-20 dB. The demonstrated channels span 1060 μm × 670 μm, and the TI-PLM chip with 1358×800 micromirrors within a 14.7 ×8.6 $mm^2$ area will produce ~156 such channels against ideal ~374 reconfigurable channels, limited by the non-ideal beam waist due to bulk optics. Further improvements can be made by adjusting the beam radius projected onto the mirror to the optimal value of ω=236 × 333 μm instead of the current value of ω=332 × 533 μm. Furthermore, >1,000 channels are feasible at this pixel pitch by scaling up the number of mirrors towards the reticle limit.

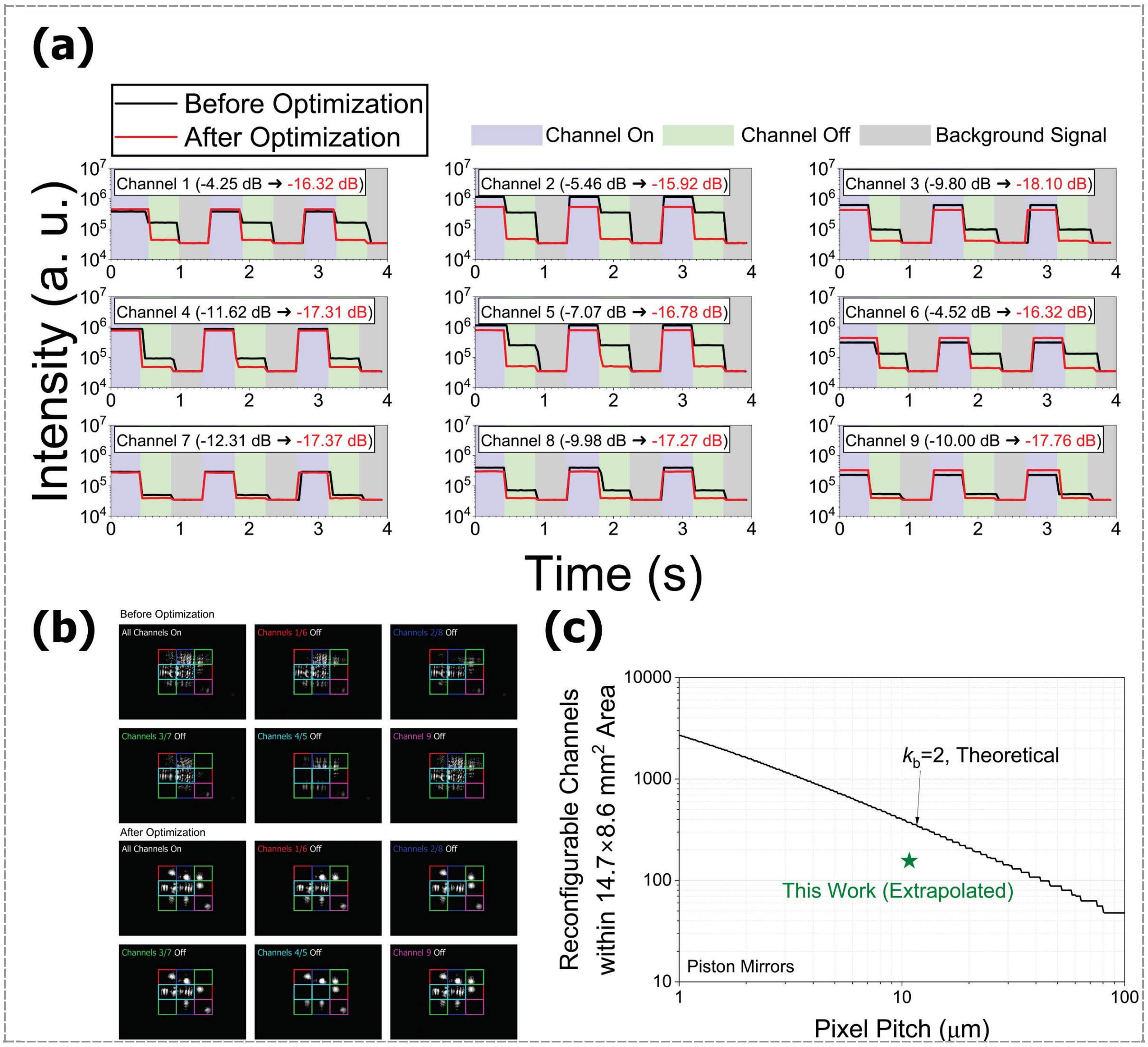

**Fig. 5 | Optimization during Simultaneous Illumination.** (a) The optimization of nine channels during simultaneous illumination, enabled by modulating different channel pairs and extracting the crosstalk from other channels. On average, the OFF/ON ratio decreased from -8.33 dB to -17.02 dB. (b) Camera image showing the modulation of each channel before and after the optimization. (c) Theoretical channel scaling for piston-motion mirrors is again shown, and the data points obtained in this work are indicated. When extrapolated, the channel density shown here approaches the theoretical value.

Figure 6 shows an additional advantage for piston-motion mirrors: enabling one-to-many mapping, which can send the beam to two or four different positions with similar intensities. This way, one can enable a unidirectional optical fan-out. Figure 6 (a) shows the output before optimization, where patterns A–D are described in Methods. Ideally, pattern A should direct 16.4 % of the beam to each diagonal spot, while patterns B and C should direct 40.5 % of the beam to two symmetric spots. However, the output shows a number of unwanted diffraction orders. Figure 6 (b) reveals optimized worst-case crosstalk of -18.3 dB for A, -19.4 dB for B, -18.8 dB for C, and -19.6 dB for D, showing that multi-output channels have larger crosstalk. Currently, returning patterns are 180° rotated copies of patterns A–D to ensure vertical incidence to the camera, contributing to further signal reduction due to the fan-out at the second reflection. It is expected that the optimization can be improved by separate optimization of the first and second reflectors, while ensuring vertical incidence with hardware such as microlens-integrated small photodetectors.

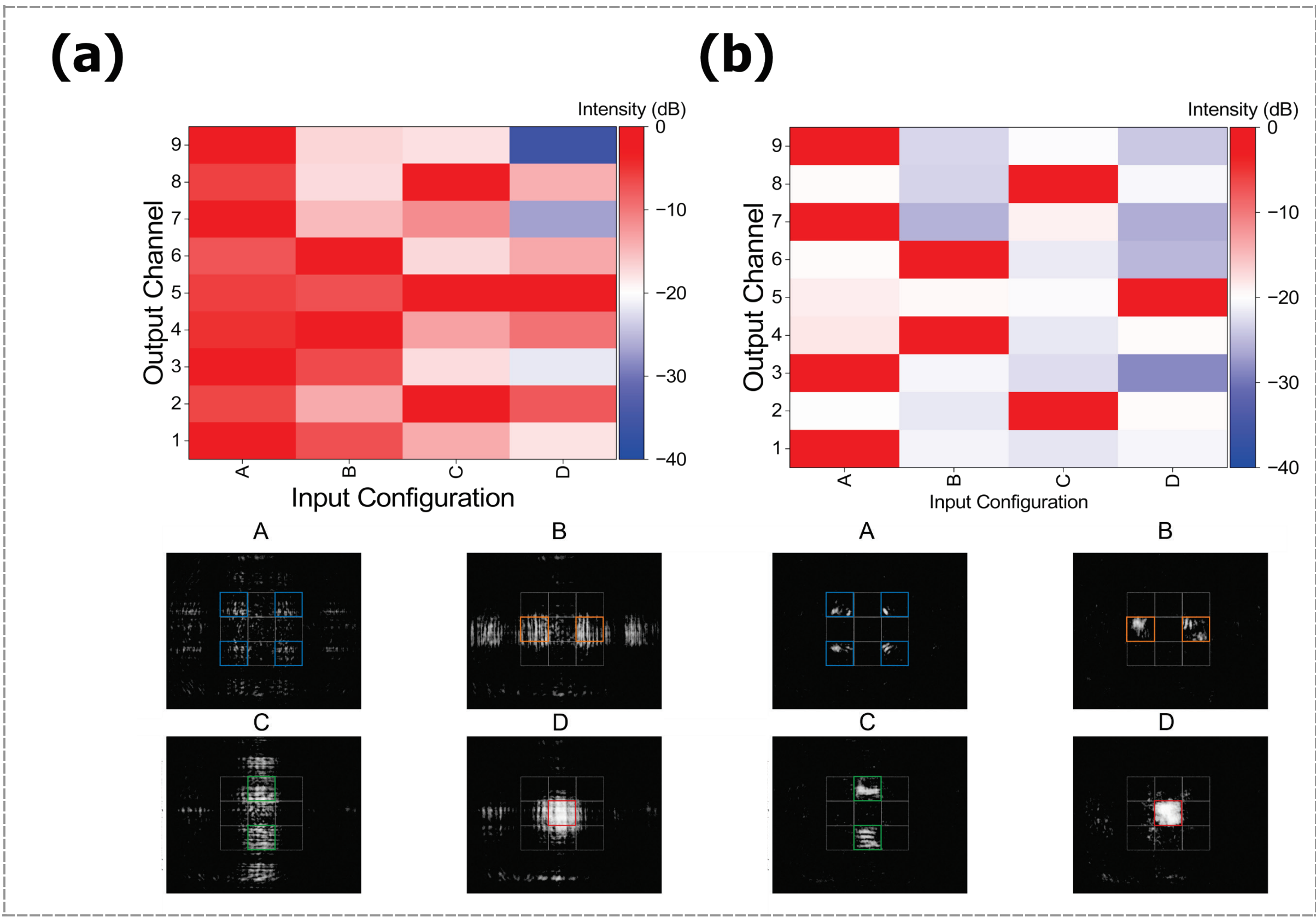

**Fig. 6 | Optimization for multi-output patterns.** Optimization heatmap and the camera image (a) before and (b) after the optimization. Multi-output patterns can also be optimized with piston-motion micromirrors, although slightly degraded performance with increasing number of outputs indicates further room for optimization improvements, mostly from increasing the signal intensity.

# Discussion

A free-space optical interconnect system based on piston-motion micromirror arrays is proposed and its reconfigurability is experimentally verified. Co-optimizing nine different configurations achieved < -20 dB crosstalk for $k_b$=2, which is impossible with flat tip-tilt-only mirrors. The method also works in the presence of other working channels, making it practical for real systems. The demonstrated channel spacing and size are expandable to ~156 reconfigurable channels on a single TI-PLM chip. Achieving ~1,000 channels at this pitch requires scaling down the aperture size while scaling up the chip to the reticle limit. Piston-motion mirrors can further enable optical fan-out with multi-output patterns, which can be particularly useful for resource disaggregation such as memory-to-graphics processing units (GPUs) connections. Overall, we demonstrate two advantages of piston-motion mirrors over tip-tilt mirrors — (i) reducing the channel crosstalk through pixel optimization, and (ii) multicasting of one input to multiple outputs.

The study raises an important question about whether tip-tilt mirrors or piston-motion mirrors are more favorable for optical interconnects. On one hand, tip-tilt mirrors may have an advantage because a large tip-tilt angle (e.g., ≥19.25°) is practically more achievable than submicron mirror pitch (e.g., ≤360 nm). For instance, the smallest academically demonstrated piston mirror pitch is 500 nm[22], while large tip-tilt angle mirrors are academically demonstrated as large as ±60° mechanical[23]. Furthermore, large channel counts require many more piston-motion mirrors than tip-tilt mirrors. For instance, 10,000 fully reconfigurable channels require 100×100 tip-tilt mirrors or 25,000×25,000 piston mirrors (250×250 mirrors per channel). Thus, yield becomes a concern for piston-motion mirrors. In addition, angle-dependent diffraction efficiency can be a problem for large-pitch mirrors.

On the other hand, high fill-factor analog tip-tilt mirrors are not readily available, while piston-motion micromirror arrays are increasingly becoming commercial with pitches such as 10.8 μm[16] or 16 μm[20].

Furthermore, pixel optimization and multicasting in this work can further reduce the channel count requirements for piston-motion mirrors. For instance, multiple communication channels can fit within a single reconfigurable channel. For micro-LED-based optical interconnects with a 3.3 Gbps/lane data rate and 50 µm pitch[5], approximately 36 micro-LEDs can fit within one channel shown in this work, resulting in ~118 Gbps/channel data rate. Within the lane, electrical reconfiguration can be combined with beam profile control/alignment from piston-motion mirrors. Although the idea may require complex optical design and thus feasibility must be tested, it suggests the advantage of piston-motion mirrors. Similarly, piston-motion mirrors can be used in low-channel-count applications such as fiber-optic switches[14], enabled by phase modulators for the near-infrared such as TI-PLM NIR[24]. Finally, the pixel optimization can adjust the focus of the beam[18]. Combined with multiple (e.g., ≥ 4) reflections, focusing between each reflection overcomes the trade-off between aperture size and separation described in Eqs. (1a-2b), as it allows the beam to have a larger travel distance with minimal diffraction.

Still, future hardware improvements, such as large-angle tip-tilt mirrors or small-pitch piston-motion mirrors, can benefit the research further. In addition, piston-motion mirrors will benefit from increased phase resolution that can minimize the scattering loss in fine optimizations. Finally, tip-tilt-piston mirrors[25] can be a unique approach, merging the benefits of both tip-tilt and piston-motion mirrors. On the other hand, non-MEMS phase-modulation approaches such as liquid crystal on metasurfaces[26] can also be useful in scaling down the pixel pitch.

# Methods

### Base Phase-pattern construction

Tip-tilt patterns in the main text show a phase gradient that is calculated from an analytic equation:

$$\varphi(x_{in}, y_{in}, \theta_x, \theta_y) = mod\left(\frac{2\pi}{\lambda}\left(\left\{\sin\left(\frac{\pi}{4}+\theta_x\right)-\sin\left(\frac{\pi}{4}\right)\right\}x_{in}+sin\theta_y y_{in}\right), 2\pi\right) \quad (5)$$

where $(x_{in}, y_{in})$ are the displacements from the pattern center, $\theta_x, \theta_y$ are optical deflection angles in the x- and y-directions, and $\lambda$ is the wavelength. The phase level is calculated at the center of each pixel and set to the closest available phase level.

The initial phase patterns used in Figure 6 are defined by:

$$\varphi_A(x_{in}, y_{in}) = (\varphi_1 - \varphi_0)mod\left(\left\lfloor\frac{x_{in}}{h\Lambda}\right\rfloor + \left\lfloor\frac{y_{in}}{h\Lambda}\right\rfloor, 2\right) + \varphi_0 \quad (6a)$$

$$\varphi_B(x_{in}, y_{in}) = (\varphi_1 - \varphi_0)mod\left(\left\lfloor\frac{x_{in}}{h\Lambda}\right\rfloor, 2\right) + \varphi_0 \quad (6b)$$

$$\varphi_C(x_{in}, y_{in}) = (\varphi_1 - \varphi_0)mod\left(\left\lfloor\frac{y_{in}}{h\Lambda}\right\rfloor, 2\right) + \varphi_0 \quad (6c)$$

$$\varphi_D(x_{in}, y_{in}) = \varphi_0 \quad (6d)$$

where $\Lambda$ is the pixel pitch, $h$ is the half-period of the pattern in pixels, $\varphi_0 = 0$ is the base level, and $\varphi_1 = 0.92\pi$ is the binary grating level. $\lfloor\ \rfloor$ denotes the floor operation.

### Optimization Method

The optimization uses gradient ascent, where the N-dimensional gradient is determined experimentally from $N^2$ different perturbations on the pattern and measurements on the region of interest from the camera. The optimization steps progress from coarse pixels (i.e., 3×3 perturbations per channel) to fine pixels (i.e., 36×36 perturbations per channel). The perturbations follow the patterns from the smallest known Hadamard matrix[27] whose order is equal to or larger than the number of elements to introduce mutually-orthogonal perturbations above the noise limit. The crosstalk is minimized by maximizing the following figure of merit:

$$J_i = 10\log_{10}\frac{I_{i,i} - I_{bg}}{\max\limits_{j\neq i} I_{i,j} - I_{bg}} \tag{7}$$

where $J_i$ is the figure of merit for the $i$ th pattern, $I_{i,j}$ is the mean intensity of the $j$ th region of interest (ROI) when the main pattern is the $i$ th pattern, and $I_{bg}$ is the mean background intensity measured from the setup. Maximizing $J_i$ is equivalent to minimizing the worst crosstalk to other cells at configuration $i$, $-J_i$. The optimization flow is designed for regular photodetector arrays, so the full camera-captured intensity profile is not used; the only information available to the optimizer is the integrated intensity of each ROI. All 9 patterns are optimized by optimizing the collective figure of merit:

$$FOM = \sum_{i=1-9}\frac{\exp(-J_i)}{\sum_{i=1-9}\exp(-J_i)}J_i \tag{8}$$

which is the soft-maximum of $J_1$,$J_2$, …, $J_9$. The softmax function makes the FOM differentiable while preserving its definition as optimizing the worst (i.e., minimum) $J_i$.

## Data availability

The data that support the findings of this study are available within the article and the supplementary material.

## Code availability

The code for the optimization is submitted along the manuscript and available upon request.

## Acknowledgements

The information, data, or work presented herein was funded in part by the Stanford Energy Postdoctoral Fellowship and the Precourt Institute for Energy, and in part by the Stanford Sustainability Accelerator.

## Author contributions

O. Solgaard and J. Lee conceived the research and O. Solgaard supervised the project. J. Lee designed the experiments. J. Lee and C. G. Valdez built the measurement setup and C. G. Valdez, A. R. Kroo, and D. A. B. Miller advised on troubleshooting. J. Lee performed the experiments and analyzed the data. J. Lee wrote the paper, with input from all authors. All authors contributed to the verification and revision.

## Additional information

**Supplementary information**: Supplementary information is provided as a separate document.

## Competing interests

The authors declare no competing interests.

## Artificial Intelligence Disclosure

Figures 1 (a-b) and 2 (a) were partially created by Blender, whose script was written by Claude Fable 5. The script was verified and the output image was screened by the authors. Claude Fable 5 and ChatGPT Astro 6.1 were used to assist with implementing the Python code used for experimental data collection. The physical models and numerical methods were designed by the authors, who reviewed, tested and validated all AI-assisted code.

## References

1. Shehabi, A. *et al.* *2024 United States Data Center Energy Usage Report*. LBNL-2001637 (2024).
2. Grutter, K., Salter, T. & Horton, T. Findings of the 2nd Photonics and Electronics Technology for Extreme Scale Computing Workgroup (REPETE): Design Challenges for Socket Level Photonic I/O. Preprint at https://doi.org/10.48550/ARXIV.2110.11759 (2021).
3. Seok, T. J., Kwon, K., Henriksson, J., Luo, J. & Wu, M. C. Wafer-scale silicon photonic switches beyond die size limit. *Optica* **6**, 490 (2019).
4. Song, S. *et al.* A 32 Gb/s/λ 256Gb/S/Fiber Half-Rate Bandpass-Filtered Clock-Forwarding DWDM Optical Link in a 3D-Stacked 7nm EIC/65nm PIC Technology. in *2026 IEEE International Solid-State Circuits Conference (ISSCC)* 396–398 (IEEE, San Francisco, CA, USA, 2026). doi:10.1109/ISSCC49663.2026.11409081.
5. Pezeshki, B. *et al.* 304 channel MicroLED based CMOS transceiver IC with aggregate 1 Tbps and sub-pJ per bit capability. in *Optical Fiber Communication Conference (OFC) 2024 (2024), paper M3A.1* M3A.1 (Optica Publishing Group, 2024). doi:10.1364/OFC.2024.M3A.1.
6. Goldman Sachs Global Investment Research. *Optical Networking: The Next Mega Trend in AI Infrastructure*. https://www.goldmansachs.com/pdfs/insights/goldman-sachs-research/optical-networking-the-next-mega-trend-in-ai-infrastructure/report.pdf (2026).
7. Daudlin, S. *et al.* Three-dimensional photonic integration for ultra-low-energy, high-bandwidth interchip data links. *Nat. Photon.* **19**, 502–509 (2025).
8. Miller, D. A. B. Attojoule Optoelectronics for Low-Energy Information Processing and Communications. *Journal of Lightwave Technology* **35**, 346–396 (2017).
9. Miller, D. A. B. Waves, modes, communications, and optics: a tutorial. *Adv. Opt. Photon.* **11**, 679 (2019).
10. Pezeshki, B. *et al.* 10-Gb/s NRZ Per Lane Data Links With Transferred Micro-LEDs With Record BER and Energy Efficiency. *IEEE Photon. Technol. Lett.* **37**, 453–456 (2025).
11. Ciftcioglu, B. *et al.* 3-D integrated heterogeneous intra-chip free-space optical interconnect. *Opt. Express* **20**, 4331 (2012).
12. Wang, K. *et al.* High-Speed Reconfigurable Free-Space Optical Interconnects with Carrierless-Amplitude-Phase Modulation and Space-Time-Block Code. *J. Lightwave Technol.* **37**, 627–633 (2019).
13. Hagelin, P. M., Krishnamoorthy, U., Heritage, J. P. & Solgaard, O. Scalable optical cross-connect switch using micromachined mirrors. *IEEE Photon. Technol. Lett.* **12**, 882–884 (2000).
14. Urata, R. *et al.* Mission Apollo: Landing Optical Circuit Switching at Datacenter Scale. Preprint at https://doi.org/10.48550/ARXIV.2208.10041 (2022).
15. Lee, J., Valdez, C. G., Kroo, A. R., Miller, D. A. B. & Solgaard, O. Free-Space Reconfigurable Optical Interconnects for the Next Generation Datacenters. in *2026 International Conference on Optical MEMS and Nanophotonics (OMN)* 1–2 (IEEE, Saint-Alexis-des-Monts, QC, Canada, 2026). doi:10.1109/OMN70392.2026.11646797.

16. Ouyang, B., Gong, S., Lawrence, T. & Hall, J. Evaluating Texas instruments Phase Light Modulator (PLM). in *Ocean Sensing and Monitoring XIV* (eds Hou, W. 'Will' & Mullen, L. J.) 18 (SPIE, Orlando, United States, 2022). doi:10.1117/12.2623938.
17. Solgaard, O. Optical MEMS Fiber Switches (Chapter 8.6). in *Photonic Microsystems* 318–326 (Springer US, Boston, MA, 2009). doi:10.1007/978-0-387-68351-5_8.
18. Lee, J. & Solgaard, O. Simulation studies on overcoming two challenges in phase-modulating micromirrors. in *Emerging Digital Micromirror Device-based Systems and Applications XVIII* (eds Lyubarsky, A. & Bridger, K.) 36 (SPIE, San Francisco, United States, 2026). doi:10.1117/12.3098273.
19. Rocha, J. C. A. *et al.* Fast and light-efficient wavefront shaping with a MEMS phase-only light modulator. *Opt. Express* **32**, 43300 (2024).
20. Payne, A. *et al.* Planar Light Valve and Displacement Phase Modulator: MEMS-based spatial light modulators for high-speed amplitude and phase modulation. *J. Optical Microsystems* **6**, (2026).
21. Magnusson, R. & Gaylord, T. K. Diffraction efficiencies of thin phase gratings with arbitrary grating shape. *J. Opt. Soc. Am.* **68**, 806 (1978).
22. Stahl, R. *et al.* Modular sub-wavelength diffractive light modulator for high-definition holographic displays. *J. Phys.: Conf. Ser.* **415**, 012057 (2013).
23. Pollock, C., Javor, J., Stange, A., Barrett, L. K. & Bishop, D. J. Extreme angle, tip-tilt MEMS micromirror enabling full hemispheric, quasi-static optical coverage. *Opt. Express* **27**, 15318 (2019).
24. Walker, Z. *et al.* Texas Instruments phase light modulator (PLM) for near infrared applications. in *Emerging Digital Micromirror Device-based Systems and Applications XVIII* (eds Lyubarsky, A. & Bridger, K.) 11 (SPIE, San Francisco, United States, 2026). doi:10.1117/12.3077906.
25. Jung, I. W., Krishnamoorthy, U. & Solgaard, O. High Fill-Factor Two-Axis Gimbaled Tip-Tilt-Piston Micromirror Array Actuated by Self-Aligned Vertical Electrostatic Combdrives. *J. Microelectromech. Syst.* **15**, 563–571 (2006).
26. Liu, X., Duan, X., Li, B., Zheng, Z. & Li, Q. Liquid Crystal on Metasurfaces Toward Dynamic Photonic Control. *Adv Funct Materials* **36**, e29188 (2026).
27. Hedayat, A. & Wallis, W. D. Hadamard Matrices and Their Applications. *Ann. Statist.* **6**, (1978).

# Supplementary Information for Reconfigurable Free-Space Optical Interconnects

Jaekwon Lee[1,2], Carson G. Valdez[1,2], Anne R. Kroo[1,2], David A. B. Miller[1,2], and Olav Solgaard[1,2,*]

[1]Edward L. Ginzton Laboratory, Stanford University, 348 Via Pueblo, Stanford, 94305, CA, USA

[2] Electrical Engineering, Stanford University, 350 Jane Stanford Way, Stanford, 94305, CA, USA

[*]Corresponding author. E-mail: solgaard@stanford.edu

## The optimization of angle of incidence for free-space optical interconnects

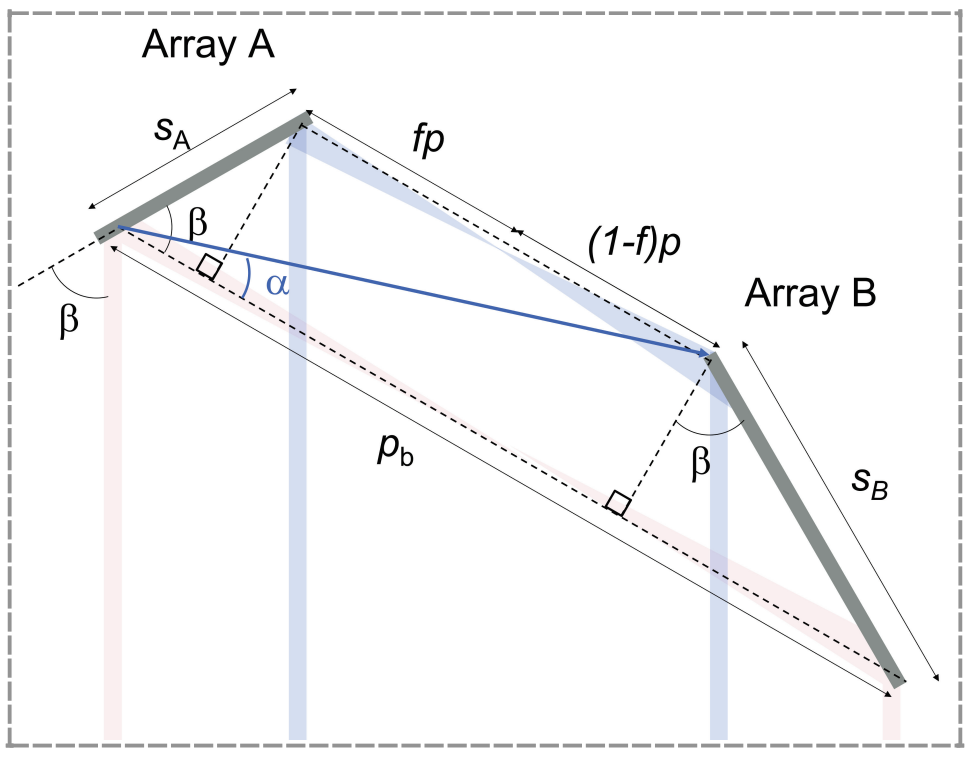

**Fig. S1 | Schematic of the Optical Interconnect system at an arbitrary angle β.**

Let us consider the optical interconnect system in Figure S1, showing two mirrors placed at the angles of β and π/2-β, respectively, in order to achieve parallel input/output beams. Here, β is assumed to be π/4≤β<π/2 without losing generality. In this case, the size of the second mirror array can be derived from connecting two parallel lines connecting each end of the array and comparing the expression for the distance between two lines, $\boldsymbol{s_B cos\beta = s_A sin\beta}$, thus:

$$s_B = s_A tan\beta \tag{S1a}$$

where $s_A$ and $s_B$ refer to the length of each mirror array, respectively. The number of channels should match between the two arrays, so Eq. (S1a) translates into:

$$d_B = d_A tan\beta \tag{S1b}$$

where $d_A$ and $d_B$ refer to the size of individual channels at each mirror array, respectively. Now, if the focal point is placed at a distance of *fp* from mirror 1 and (1-*f*)*p* from mirror 2, substituting $d_A sin\beta = k_b\omega_A$ and $d_B cos\beta = k_b\omega_B$ into Eq. (S1b) results in $\omega_A = \omega_B$ and thus:

$$\omega_0\sqrt{1+\left(\frac{\lambda fp}{\pi\omega_0^2}\right)^2} = \omega_0\sqrt{1+\left\{\frac{\lambda(1-f)p}{\pi\omega_0^2}\right\}^2} \tag{S2}$$

where $p$ is the distance between the mirror arrays, $\omega_0$ is the beam radius at the focal point, and λ is the wavelength. Solving Eq. (S2) gives $f = 0.5$ for any angle β, so the focal point always needs to be placed at an equal distance from both arrays in order to maximize the number of channels.

As π/4≤β<π/2, the channel count is limited by the case where the bottom of array 1 is connected to the top of array 2, and the equation for the channel count can be written as[1]:

$$N = 1 + \frac{\{p_b - (N-1)(d_A cos\beta + d_B sin\beta)\}tan\alpha}{d_A(sin\beta - cos\beta tan\alpha)} \tag{S3a}$$

where $N$ is the linear channel count that is the same for both arrays, and $p_b$ is the separation at the bottom of the interconnect. Organizing Eq. (S3a) gives:

$$N = 1 + \frac{p_b tan\alpha}{k_b \omega_A (1 + tan\alpha tan\beta)} \leq 1 + \frac{\sqrt{\pi p_b} tan\alpha}{k_b \sqrt{\lambda}(1 + tan\alpha)} \tag{S3b}$$

where the equality holds when $\omega_0 = \sqrt{p\lambda/2\pi}$ (i.e., $\omega_A = \sqrt{2}\omega_0$ for any β) and tan β=1 (β=π/4). Therefore, β=π/4 maximizes the number of channels in the proposed system.

## Derivation of Scaling Equations for free-space optical interconnects

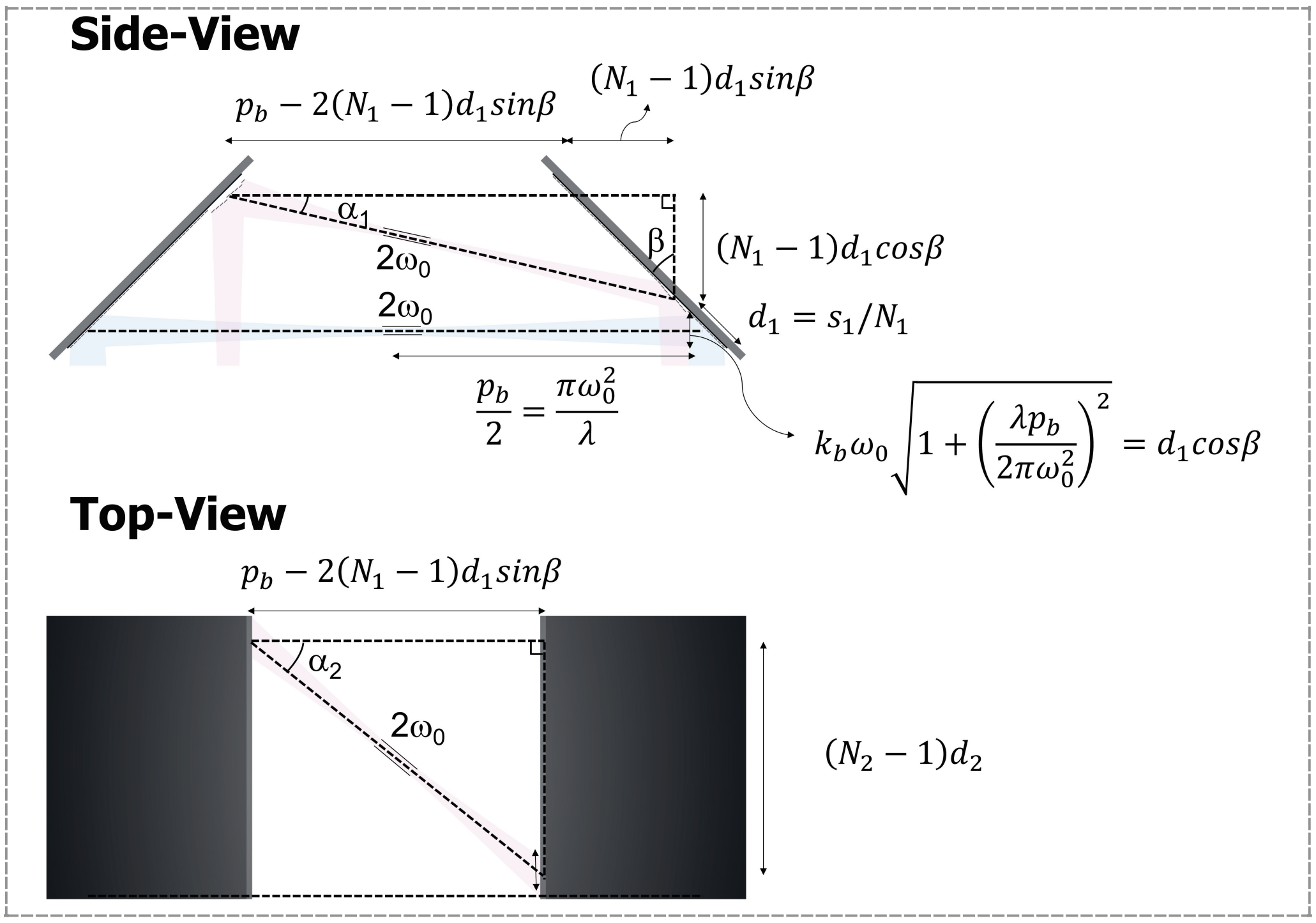

**Fig. S2 | Schematic of the Optical Interconnect system at β=45°.**

To maximize the reconfiurable channel count, we balance two conflicting conditions: (1) minimizing the aperture on the mirror array to ensure maximum channel density, and (2) maximizing the separation between the two micromirror arrays so that a given maximum deflection angle can have enough offset in order to maximize the number of output mirrors each channel can reach. The trade-off between aperture size and micromirror separation is governed by diffraction, and we will find optimal $\omega_0$ at a given set of system parameters.

Now, given that β=π/4 as shown above, Figure S2 can be used for the derivation of the scaling equation. The derivation starts with the tilted edge of the chip, utilizing the fact that the number of reconfigurable channels that are reachable from the end of one array follows the equation[1]:

$$tan\alpha_1 = \frac{(N_1 - 1)d_1 cos\beta}{p_b - (N_1 - 1)d_1 sin\beta} \tag{S4a}$$

where $N_1$ is the linear channel count along the tilted edge, $d_1$ is the aperture size along the tilted edge, $\alpha_1$ is the maximum optical deflection angle along the tilted edge, β is the incidence angle of the beam, and $p_b$ is the longest linear separation between the two arrays. Utilizing the small-angle approximation (i.e., $tan\alpha_1 \approx \alpha_1$ and $cos\beta - sin\beta tan\alpha_1 \approx cos\beta$) and constant-separation approximation (i.e., $p_b \gg (N_1 - 1)d_1 sin\beta$), Eq. (S4a) reduces to $N_1 \approx 1 + p_b\alpha_1/d_1 cos\beta$, which is used in fiber-optic switches[1]. Given cosβ ≈ sinβ at β≈π/4, Eq. (S4a) can be reorganized without small-angle approximation and constant-separation assumption:

$$N_1 = 1 + \frac{p_b tan\alpha_1}{\mathrm{d}_1 \cos\beta(1 + tan\alpha_1)} \tag{S4b}$$

where $\alpha_1$ is replaced by $tan\alpha_1/(1+tan\alpha_1)$ from the equation for fiber-optic switches to remove the small-angle and constant-separation approximations. At $\alpha_1=\pi/4$, Eq. (S4b) implies $p_b = 2(N_1 - 1)d_1 \cos\beta$, which leaves the top-side separation as 0 (i.e., unphysical). Thus, the valid range of the equation is $0<\alpha_1<\pi/4$, which is expected because $\alpha=\pi/4$ implies that the deflected beam is parallel to the output mirror surface. It is worth noting that the beam steering is evaluated for the shortest unsteered beam path, because it is the limiting factor for the number of reconfigurable channels.

For the proposed system, the optimal beam waist is placed equally from the two mirrors, as shown in the previous section (i.e., $f$=0.5). In this case, the following equation holds from Figure S2:

$$d_1 cos\beta = k_b\omega_0\sqrt{1+\left(\frac{\lambda p_b}{2\pi\omega_0^2}\right)^2} \tag{S5}$$

where $\lambda$ is the wavelength, $\omega_0$ is the Gaussian beam radius at the beam waist, and $k_b$ is an application-dependent parameter indicating channel spacing ($k_b \geq 2$). Substituting Eq. (S5) into Eq. (S4b) gives:

$$N_1 = 1 + \frac{p_b tan\alpha_1}{k_b\omega_0\sqrt{1+\left(\frac{\lambda p_b}{2\pi\omega_0^2}\right)^2}(1+tan\alpha_1)} \tag{S6}$$

and the optimal $\omega_0$ can be derived from $dN_1/d\omega_0 = 0$, resulting in the optimum condition $p_b/2 = \pi\omega_0^2/\lambda$ and $d_1 cos\beta = \sqrt{2}k_b\omega_0$, meaning that each mirror is exactly one Rayleigh length away from the beam waist. It is worth noting that $p_b$ is used in this equation because it is the longest unsteered beam path and thus the limiting factor for the diffraction.

Combining the optimal $\omega_0$ from the longest beam path and the channel count scaling for the shortest beam path, the constant-separation approximation can be removed:

$$N_1 = 1 + \frac{\pi d_1 cos\beta}{\lambda k_b^2}\frac{tan\alpha_1}{1+tan\alpha_1} \tag{S7}$$

Thus, if we assume $s_1 = N_1 d_1$ as the minimum linear dimension of the micromirror array along the tilted edge, the dimension follows the equation:

$$s_1 = \frac{N_1(N_1-1)\lambda k_b^2}{\pi cos\beta}\frac{1+tan\alpha_1}{tan\alpha_1} \tag{S8}$$

For the other edge that is not tilted, the derivation remains similar starting from the equation:

$$tan\alpha_2 = \frac{(N_2-1)d_2}{p_b - 2(N_1-1)d_1 sin\beta} \tag{S9}$$

where $N_2$, $d_2$, and $\alpha_2$ are the linear channel count, the aperture size, and the maximum deflection angle along the other edge, respectively. It is worth noting that the denominator is still a function of $N_1$ and $d_1$ because we are evaluating the beam deflection in the other edge at the top of the array as well. Following a similar derivation, it can be shown that:

$$s_2 = \frac{N_2(N_2-1)\lambda k_b^2}{\pi tan\alpha_2}\frac{1}{1-\frac{2(N_1-1)d_1 sin\beta}{p_b}} \tag{S10}$$

where $s_2 = N_2 d_2$ is the minimum linear dimension of the micromirror array along the other edge. Similar to the tilted axis, $\alpha_2$ is replaced by $tan\alpha_2$ to remove the small-angle approximation, and a factor of $1/\{1-2(N_1-1)d_1 sin\beta/p_b\}$ is added to remove the constant-separation approximation.

## Calibration of phase levels at a 45° angle of incidence

The calibrartion is achieved by changing the amount of voltage applied to all mirrors at each phase level, controlling the stroke of the mirrors[2]. For this one-time calibration before the system is fully installed, the camera is placed at the position of the fixed mirror and the intensities of the 0th order and the 1st order of binary grating patterns are used to calibrate the phase levels following the binary grating equation[3]:

$$\frac{I_1}{I_0} = \frac{A\sin^2\left(\frac{\psi}{2}\right) + I_{bg}}{\cos^2\left(\frac{\psi}{2}\right) + I_{bg}} \approx A\tan^2\left(\frac{\psi}{2}\right) \tag{S11a}$$

where $I_0$ and $I_1$ are the intensities at the 0th and 1st diffraction order, ψ is the phase information under a given bias condition, $I_{bg}$ is the background intensity, and $A$ is the experimental constant defining the power ratio between the 0th and 1st diffraction order, ideally $A = 4/\pi^2$. In order to compensate for non-idealities, the measured relationship between $I_1$ and $I_0$ was fitted to the following equation:

$$I_1 = A\sin^2\left(\frac{\psi}{2}\right) + I_{bg} = A(1 - I_0 + I_{bg}) + I_{bg} \tag{S11b}$$

to determine the constant $A$ from the fitted slope instead of the analytic $4/\pi^2$. In our setup, A was 1.267. Then, the phase levels at each bias voltage is extracted using this A value from Eq. (S11a). Figure S3 (a) shows all 16 phase levels as a function of calibration voltage (i.e., mirror bias) following the procedure described in the main text. It can be seen that as the maximum bias between the mirror and the actuator plate increases (i.e., calibration voltage decreases), the phase of each level also increases. When the calibration voltage is below 0.58V, the phase for phase level 15 exceeds 2π and wraps into the lower phase levels. Considering the wrapping, the maximum voltage was chosen so that the maximum difference between nearby phase levels is minimized[4]. This occurred at a calibration voltage of 0.87V, at which the maximum difference is the difference between phase levels 14 and 15, approximately 0.3π. Figure S3 (b) shows all phase levels at the calibration voltage of 0.87 V. The inset of Figure S3 (b) lists these data points and the measured intensities overlaid on the theoretical cosine and sine curves illustrated in Eq. (S11a). It can be seen that the extracted phase levels agree well with the theoretical curves, where the fifteenth state produced a phase shift of 1.76π. Due to the strong non-linearity induced by electrostatic force in MEMS mirrors, the error between the ideal and displayed phases typically gets larger towards 2π phase shift.

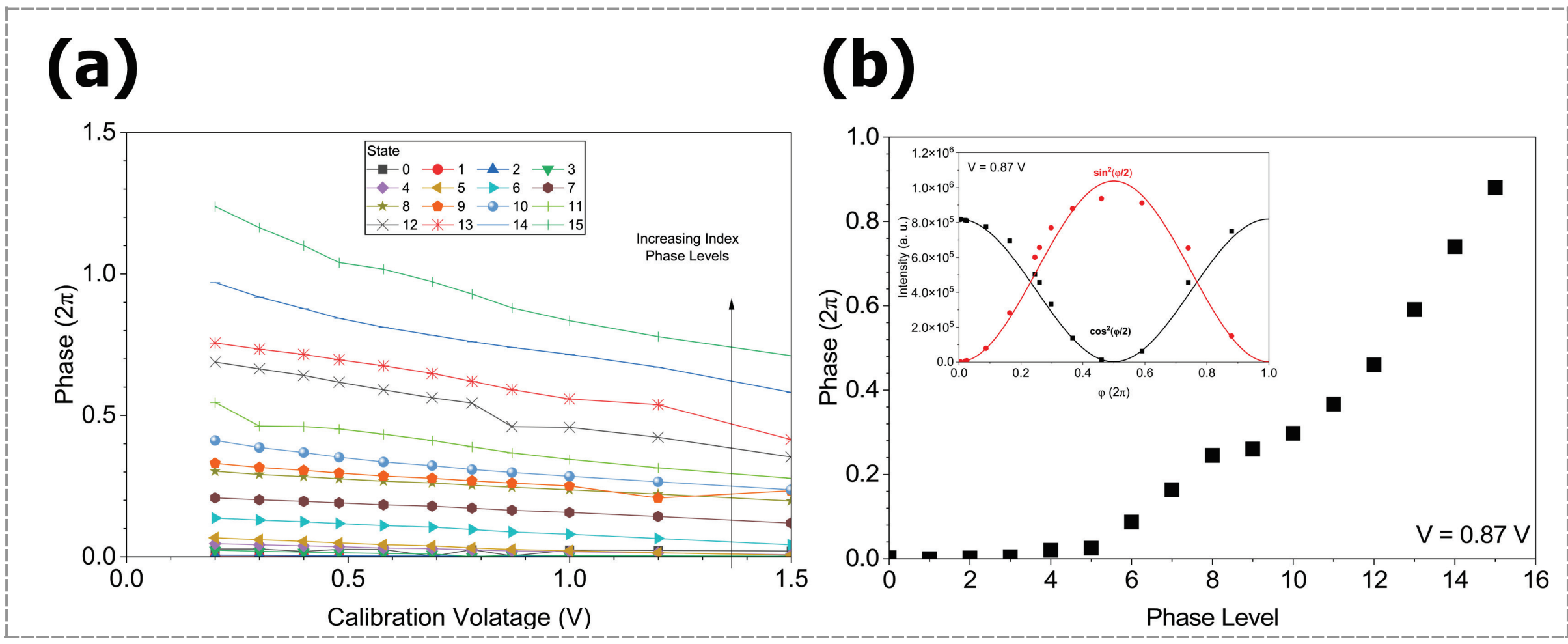

**Fig. S3 | Phase level calibration for a 45° incidence angle.** The optimum phase level distribution was found by measuring each phase level at a specific calibration voltage and minimizing the maximum difference between states.

## Proof-of-concept Simulations for Multimode Sources

To evaluate the feasibility of the concept for incoherent emissions such as multimode VCSELs or micro-LEDs, a simulation is performed with a source that has 30 nm FWHM Gaussian spectral distribution with a center wavelength at 450 nm and has 6 spatial modes, Laguerre-Gaussian modes $LG_{pl}$ with $2p+|l|+1 \leq 3$. The power distributed to each spatial mode is 34 % $LG_{00}$, 16.5 % $LG_{01}$, 16.5 % $LG_{0-1}$, 11 % $LG_{10}$, 11 % $LG_{02}$, and 11 % $LG_{0-2}$, respectively. The spatial modes are assumed to be mutually incoherent, and their propagated intensities are summed with modal power weights. The spectral response was calculated by evaluating the response at each wavelength at 5 nm intervals across 23 wavelengths. Assuming the phase levels are represented with respect to 450 nm, the effect of phase levels on each wavelength is scaled by the wavelength ratio with respect to 450 nm. The optimization for the multimode source is compared with the single-mode source case, where a fundamental Gaussian mode with monochromatic wavelength at 450 nm is used. Each source is assumed to have $\omega_0$=25 μm beam radius and thus each mirror channel is assumed to be 100 μm × 70.7 μm, which is rectangular due to the 45 degree of incidence along the X direction. The phase levels of each PLM are assumed to be continuous (i.e., unlike discrete levels in TI-PLM) and the PLM pixels are assumed to be 1 μm with a 100 % fill factor to test the ideal case. Meshes with 250 nm resolution on the PLM plane are used for the simulation, and the simulation fields from each bounce from the mirror are calculated from the angular spectrum method (ASM):

$$E(X_{out}, Y_{out}; \lambda) = \mathcal{F}^{-1}\left[\mathcal{F}\{e^{j\varphi(x_{in}, y_{in}; \lambda)}E_{in}(x_{in}, y_{in}; \lambda)\}e^{jkz\sqrt{1-(\lambda f_x)^2-(\lambda f_y)^2}}\right] \tag{S12}$$

where $\mathcal{F}$ and $\mathcal{F}^1$ denote the Fourier Transform and the inverse Fourier transform, respectively, $f_x$ and $f_y$ are transverse spatial frequencies, $z$ is the distance between the mirror and the imaging plane, $k$ is the wavenumber, $\lambda$ is the wavelength, $\varphi(x_{in},y_{in})$ is the phase shift due to the mirror displacement, $E_{in}(x_{in},y_{in};\lambda)$ is the input electric field, and $E_{out}(X_{out},Y_{out};\lambda)$ is the output electric field at the given wavelength. At the detector plane, it is assumed that each detector region of interest accepts light within ±2° of incidence angle with respect to the normal incidence. Angular filtering is applied after the entrance aperture to implement this behavior.

Figure S4 shows the optimization simulation. Similar to Figure 5 (a)-(b), the OFF/ON ratio of each channel is collectively optimized by calculating the gradient using the adjoint method. The channel mapping can be arbitrary as the simulation now assumes two PLMs. Thus, the channel mapping of channel 1 to 6, 2 to 8, 3 to 7, 4 to 5, 5 to 1, 6 to 3, 7 to 2, 8 to 4, and 9 to 9 was chosen for demonstration. The results reveal that the mirror optimization can correct for both the multi-moded nature of the beam and the spectral width of the beam, although its optimized performance remains worse than that of the monochromatic fundamental Gaussian source.

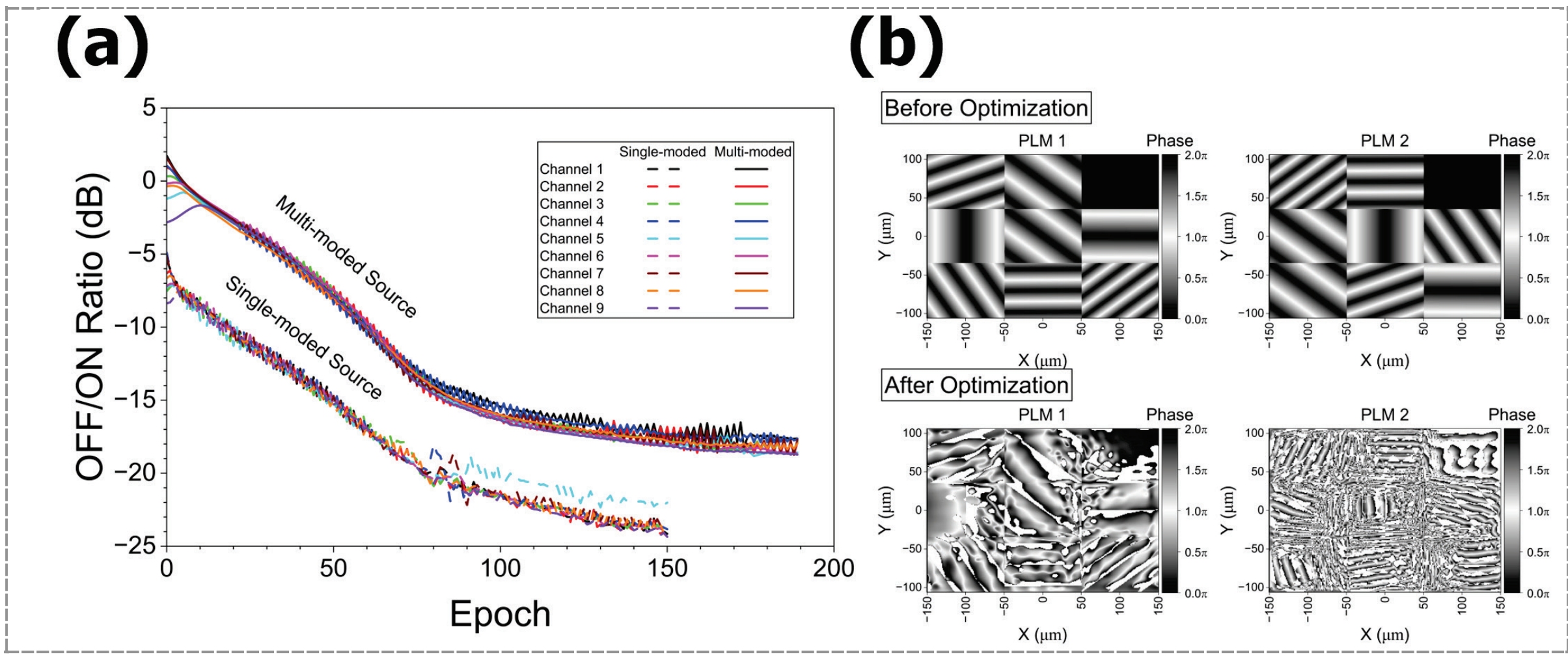

**Fig. S4 | Optimization of the multimode source.** (a) The OFF/ON ratio optimization for each channel is shown for the multimode source, and the optimization of the single-mode monochromatic source is shown as a comparison. (b) The mirror pattern before and after the optimization.

# References

1. Solgaard, O. Optical MEMS Fiber Switches (Chapter 8.6). in *Photonic Microsystems* 318–326 (Springer US, Boston, MA, 2009). doi:10.1007/978-0-387-68351-5_8.
2. Ouyang, B., Gong, S., Lawrence, T. & Hall, J. Evaluating Texas instruments Phase Light Modulator (PLM). in *Ocean Sensing and Monitoring XIV* (eds Hou, W. 'Will' & Mullen, L. J.) 18 (SPIE, Orlando, United States, 2022). doi:10.1117/12.2623938.
3. Magnusson, R. & Gaylord, T. K. Diffraction efficiencies of thin phase gratings with arbitrary grating shape. *J. Opt. Soc. Am.* **68**, 806 (1978).
4. Lee, J. & Solgaard, O. Simulation studies on overcoming two challenges in phase-modulating micromirrors. in *Emerging Digital Micromirror Device-based Systems and Applications XVIII* (eds Lyubarsky, A. & Bridger, K.) 36 (SPIE, San Francisco, United States, 2026). doi:10.1117/12.3098273.